# Mn$_2$C monolayer: hydrogenation/oxygenation induced strong room-temperature ferromagnetism and potential applications

Xiaoming Zhang,[a,b] Tingli He,[a,b] Weizhen Meng,[a,b] Lei Jin,[a,b] Ying Li,[b] Xuefang Dai,[b] and Guodong Liu[a,b,*]

[a]State Key Laboratory of Reliability and Intelligence of Electrical Equipment, Hebei University of Technology, Tianjin 300130, China.
[b]School of Materials Science and Engineering, Hebei University of Technology, Tianjin 300130, China.
E-mail: gdliu1978@126.com

**Abstract:**

Two-dimensional ferromagnetic materials with strong ferromagnetism and high Curie temperature are significantly desired for the applications of nanoscale devices. Here, based on first-principles computations, we report hydrogenated/oxygenated Mn$_2$C monolayer is a such material with strong room-temperature ferromagnetism. The bare Mn$_2$C monolayer is an antiferromagnetic metal with the local magnetic moment of Mn ~ 3$\mu_B$. However, the antiferromagnetic coupling of Mn atoms can transform into the ferromagnetic order under hydrogenation/oxygenation. Especially, the magnetic moments in hydrogenated/oxygenated Mn$_2$C monolayer can be as large as 6 $\mu_B$ per unit cell, and the Curie temperatures are above 290K. Beside the potential applications in spintronic devices, our work suggests that Mn$_2$C monolayer is also promising to be used in hydrogen/oxygen detection and removal devices.

## 1 Introduction

Since the discovery of graphene, a large number of two dimensional (2D) materials including transition-metal dichalcogenides, silicene, transition metal carbide, nitride, and carboniborides (known as MXenes), and borophene have been synthesized.[1-11] They exhibit many exotic properties and have a wide range of practical applications such as electronic and spintronic devices.[12-18] However, their applications in spintronics are greatly limited currently, because most known 2D materials are intrinsically nonmagnetic (NM). Therefore, developing 2D materials with room-temperature ferromagnetism is highly desired. Recently, the experimental realization of ferromagnetism in 2D $Cr_2Ge_2Te_6$, $CrI_3$ and $Fe_3GeTe_2$ has generated intensive interests.[19-21] Beside exploring 2D materials with intrinsic ferromagnetism, several approaches such as strain engineering, doping, defect generation, and surface passivation have been developed to introduce magnetism in 2D materials. For example, spin polarization will be established in layered $NbS_2$ and $NbSe_2$ by applying biaxial tensile strain;[22] graphene and phosphorene can exhibit weak magnetism under elemental doping;[23,24] $MoS_2$ monolayer can also become ferromagnetic (FM) under the combined effects of strain and hydrogenation.[25] However, the magnetism in these systems is usually quite weak, which still limits their practical applications in spintronic devices.

Fortunately, it has been demonstrated recently that MXenes can provide an excellent platform to realize strong ferromagnetism. For example, $Cr_2C$ monolayer intrinsically shows ferromagnetism with the magnetic moment as high as 8 $\mu_B$ per unit cell;[26] $Cr_2N$ monolayer can transform from antiferromagnetic (AFM) to FM state under oxygenation, and the magnetic moment can reach 9 $\mu_B$ per unit cell;[27] some other MXenes including $Ti_2C$, $Mn_2N$, and $Mn_2N$ can also exhibit strong ferromagnetism with magnetic moment of 6 - 9$\mu_B$ per unit cell when their surfaces are terminated by F, OH, Cl, or other atoms.[28-30]

Quite recently, a new Mn-based MXene, namely $Mn_2C$, is identified.[31] In $Mn_2C$ monolayer, it has seen a local magnetic moment of ~3$\mu_B$ on Mn atom. However, the spin polarization among neighboring Mn atoms naturally takes AFM ordering rather than FM ordering.[31] It is well known that, hydrogenation/oxygenation, which usually happens during the synthesis of MXene, can be an effective approach on turning the magnetic ordering in MXene.[27-30] Therefore, in this work we theoretically investigate the feasibility of inducing ferromagnetism in $Mn_2C$ monolayer by hydrogenation/oxygenation. The hydrogenation/oxygenation process on $Mn_2C$ monolayer is systematically studied, and it is shown that the fully hydrogenated/oxygenated $Mn_2C$ monolayer ($Mn_2CH_2$/$Mn_2CO_2$) exhibits a stable ferromagnetic ground state, with the total magnetic moment as large as 6 $\mu_B$ per unit cell and the Curie temperature higher than 290K. The room temperature ferromagnetism in hydrogenated/oxygenated $Mn_2C$ monolayer makes it promising for a wide range of applications such as spintronic devices, and hydrogen/oxygen (H/O) detection and removal devices.

## 2   Computational methods

In this work, the computations are performed based on density functional theory (DFT), realized by the Vienna ab initio simulation package (VASP).[32] The generalized gradient approximation (GGA) with Perdew−Burke−Ernzerhof (PBE) functional is used for exchange and correlation contributions.[33,34] During the computations, the DFT-D2 method is adopted to describe the long-range van der Waals interactions.[35] The cutoff energy is selected as 450 eV. For the structure of bare $Mn_2C$ monolayer, a vacuum spacing of more than 20 Å is built to eliminate possible interactions between neighboring layers. Considering the strong correlation effects in the Mn element, the GGA+U method is used in the calculations.[36,37] Herein, the effective Coulomb energy $U_{eff}$ of Mn atom is set as 4.0 eV. To sample the Brillouin zone, the Γ-centered $k$-mesh of 7×7×1 and 15 × 15 × 1 are adopt for structure optimization and total energy calculation. The convergence criteria for energy and force are set as $10^{-7}$ eV and 0.01 eV Å$^{-1}$, respectively.

# 3  Results and discussions

### 3.1 The ground state of Mn$_2$C monolayer

MXenes can crystallize in different structure configurations such as 2H, 1T, and 1T'. We first investigate the most energetically stable structure for Mn$_2$C monolayer. Our calculations find the 2H structure has the lowest energy among these structures, which is consistent with former computations.[31] As shown in Fig. 1(a), the 2H-type Mn$_2$C monolayer manifests a honeycomb lattice with the $D_{3h}$ symmetry, In the side view, it exhibits a triple-layer configuration with atomic layers stacked in the sequence of Mn−C−Mn, as shown in Fig. 1(b). The optimized lattice constants of Mn$_2$C monolayer are a = b=2.565 Å, with the Mn–C interatomic distance of 2.098 Å.

Then we construct a 2 × 2 supercell of Mn$_2$C monolayer to investigate its ground magnetic state. Herein, different magnetic states including NM, FM, and AFM are considered. According to the crystal symmetry, there exist three possible AFM coupling configurations (AFM1, AFM2, and AFM3, see Fig. 1c) in Mn$_2$C monolayer. The total energies for different magnetic states are shown in Fig. 1(d). We can observe that the AFM2 state possesses the lowest energy, indicating the intrinsic AFM ordering in Mn$_2$C monolayer. Our results are consistent with previous computations.[31] Under the ground state, the total and projected density of states (DOS) for Mn$_2$C monolayer is shown in Fig. 2 (e). It obviously manifests a metallic electronic structure with sizable DOS at the Fermi level. Given by the projected DOS, the states near the Fermi level are mostly contributed by the 3$d$ orbitals of Mn atoms.

### 3.2  Hydrogenation/oxygenation induced AFM-FM transition

To investigate the hydrogenation/oxygenation process on Mn$_2$C, we need first determine the most favorable adsorption site for H/O atom. We use a 4 × 4 supercell of Mn$_2$C monolayer as the substrate and first assign an isolate H/O atom on its surface. Considering the crystal symmetry, there exist four typical H/O adsorption sites on Mn$_2$C monolayer. These adsorption sites are shown in Fig. 1(a), where site-A, site-B, site-C and site-D are on the top of Mn atoms, C atoms, the bridge site of Mn-C atoms and the center of the hexagonal lattice, respectively. The adsorption energy ($E_{Ad}$) for different sites are calculated, where the $E_{Ad}$ is defined as:

$$E_{Ad} = E_{H/O+Mn_2C} - E_{Mn_2C} - E_{H/O} \qquad (1)$$

In equation (1), $E_{H/O+Mn_2C}$ and $E_{Mn_2C}$ represent the total energies of Mn$_2$C supercell after and before H/O adsorptions, and $E_{H/O}$ is the energy for per H/O atom. Our calculations show site-D possesses the lowest energy among the four adsorption sites, indicating both H and O atoms tend to be adsorbed above the center of the Mn$_2$C honeycomb lattice. Then we assign one more H/O atom on Mn$_2$C monolayer. Here two possibilities need to be considered: (1) two H/O atoms situate on the same side and (2) opposite sides of Mn$_2$C surfaces. We find the latter case is lower in energy by 65 meV for per H atom and 39 meV for per O atom. Therefore, H/O atom prefers to be adsorbed on different sides of Mn$_2$C surfaces when more atoms are included.

We continue to increase the numbers of H/O atoms on both sides of Mn$_2$C monolayer. This process well describes the increase of the hydrogenation/oxygenation degree. During this process, we compare the energy for all the possible magnetic states (NM, FM, AFM1, AFM2, and AFM3) of Mn$_2$C monolayer. In Figs. 2(a) and (b), we show the energy difference among these magnetic states under different hydrogenation and oxygenation degree, respectively. Here, a 0% hydrogenation/oxygenation degree is for the bare Mn$_2$C monolayer without H/O adsorption, and a 100% degree means fully hydrogenation/oxygenation with all the sites-D of Mn$_2$C supercell occupied by H/O atoms. In Figs. 2(a) and (b), the energy for the AFM2 state has been set as the reference energy, and the energy for the NM state is not shown in the figures because it possesses far higher energy than other magnetic states. For the hydrogenation case, as shown in Fig. 2 (a), we can observe that the energy difference between AFM2 and other magnetic states become smaller with increasing the hydrogenation degree. Especially, the energy of FM state becomes the lowest under the 100% hydrogenation degree, indicating the occurrence of AFM-to-FM transition during the period. Similar AFM-to-FM transition also happens under 75% and 100% oxygenation degree, as shown in Fig. 2 (b). Just as our expected, $Mn_2C$ monolayer can show strong ferromagnetism under proper hydrogenation/oxygenation, where the magnetic moment is 6.44 $\mu_B$ per unit cell under 100% hydrogenation degree, and 6.20 $\mu_B$ and 6.12 $\mu_B$ under 75% and 100% oxygenation degree. These values are significantly higher than previous induced ferromagnetism in graphene, phosphorene, $NbS_2$ and $MoS_2$.[22-25, 38-40]

In the following, we use the cases of 100% hydrogenation and oxygenation degree for more detailed discussions. Figure 3(a) shows the structural geometry for these states, where H/O atoms are chemically bonding to Mn atoms. Under hydrogenation and oxygenation, as shown in Fig. 3(b) and (c), we can find that FM $Mn_2C$ monolayer exhibits metallic electronic structures, where exist sizable DOS at the Fermi level in both spins. This ensures good conductivity for the applications in electronic devices. Then we investigate the stability of FM state in hydrogenated/oxygenated $Mn_2C$ monolayer by evaluating the Stoner criterion:[41]

$$I * D(E_F) > 1 \qquad (2)$$

where $I$ is the effective Stoner exchange parameter, and $D(E_F)$ is the total DOS at the Fermi level for the NM state. If the value of $I * D(E_F)$ is larger than 1, FM state is more likely to form to lower the total energy. The exchange parameter $I$ can be calculated by using the following equation:

$$I = \Delta E / \mu \qquad (3)$$

where $\Delta E$ is the exchange splitting between the bands in spin-up and spin-down, and $\mu$ is the local magnetic moment on the atom. Our calculations show the value of $I$ is similar for $Mn_2C$ monolayer before and after hydrogenation/oxygenation, which is calculated to be in the range of 0.72-0.76. Therefore, the $D(E_F)$ plays the decisive role in the stability of FM state. Figure 3(d) compare the total DOS of $Mn_2C$ monolayer under the NM state before and after hydrogenation/oxygenation. One can find that the $D(E_F)$ in hydrogenated/oxygenated Mn$_2$C is more than two times of that in pristine Mn$_2$C monolayer. Especially, both cases of hydrogenated and oxygenated Mn$_2$C satisfy $I*D(E_F) > 1$, indicating the FM state is more stable in these systems.

The Curie temperature (T$_C$) is an crucial parameter for FM materials. Here, by using the Monte Carlo (MC) simulation based on Ising model, we evaluate the T$_C$ of hydrogenated/oxygenated Mn$_2$C monolayer. The Hamiltonian for Ising model can be described as:

$$H = -\Sigma_{ij} J_{ij} M_i M_j \qquad (4)$$

where $J_{ij}$ represents the nearest-neighboring exchange parameter, and $M$ is the spin magnetic moment on Mn atom. Based on the energy difference between AFM and FM states, the exchange parameter $J$ is estimated to be 6.8 meV and 7.6 meV for hydrogenated and oxygenated Mn$_2$C monolayer, respectively. In our MC simulation, a 100 × 100 size of supercell is used, which is large enough to ensure credible results. As shown in Figs. 4(a) and (b), the estimated Curie temperature is 293K for hydrogenated Mn$_2$C and 323K for oxygenated Mn$_2$C. This indicates hydrogenated/oxygenated Mn$_2$C monolayer is promising to be used in spintronic and electronic devices working in room temperature.

### 3.3 Possible applications

Beside potential applications in spintronic devices, the novel magnetic properties make Mn$_2$C monolayer also promising to be used in H/O detection and removal devices. Figures 5(a) and (b) simply show the schematic diagrams for a H/O detection device based on Mn$_2$C monolayer. Under none or a low H/O atmosphere, Mn$_2$C monolayer is in the AFM state. The power indictor does not work because the circuit is open [see Fig. 5(a)]. However, when exposed in a high H/O atmosphere, Mn$_2$C monolayer will transfer into the FM state after hydrogenation/oxygenation. As a result, the magnet and Mn$_2$C monolayer are connected and lights the power indicator [see Fig. 5(b)].

Mn$_2$C monolayer is also promising be applied to H/O removal devices. Especially, comparing with traditional H/O removal devices, the removal devices based on Mn$_2$C monolayer possess a major advantage. Figure 6(a) shows the mechanism for traditional H/O removal, where the adsorbent is displayed for H/O adsorption. For this case, it is difficult to monitor the H/O adsorption process on the adsorbent. However, this problem can be simply solved in the Mn$_2$C-based H/O removal devices. As shown in Fig. 6(b), we display a magnet in the cavity containing H/O atmosphere. When Mn$_2$C monolayer reaches the maximum H/O adsorption capacity, it becomes ferromagnetic and will be caught by the magnet. Then an automatic H/O removal process can be realized by inputting the bare Mn$_2$C monolayer and outputting the hydrogenated/oxygenated Mn$_2$C monolayer from the cavity.

## 4  Summary

In summary, based on first-principles calculations, we investigate the feasibility of inducing ferromagnetism in Mn$_2$C monolayer by hydrogenation/oxygenation. Our calculations show both H and O can be stably bonded to the Mn atom in Mn$_2$C, and the hydrogenation/oxygenation is systematically studied. We find the antiferromagnetic Mn$_2$C monolayer becomes ferromagnetic under fully hydrogenation/oxygenation, which can be well described by the Stoner criterion. The hydrogenated/oxygenated Mn$_2$C monolayer shows strong ferromagnetism with magnetic moments higher than 6 $\mu_B$ per unit cell. By using Monte Carlo simulations, the Curie temperature is calculated to be 293K and for 323K for hydrogenated and oxygenated Mn$_2$C monolayer, respectively. The novel magnetic properties in Mn$_2$C monolayer make it promising for variable applications ranging from spintronic devices, to hydrogen/oxygen detection and removal devices.

## Acknowledgments

This work is supported by the 333 Talent Project of Hebei Province (No. A2017002020), the Special Foundation for Theoretical Physics Research Program of China (Grant No. 11747152). One of the authors (X.M. Zhang) acknowledges the financial support from Young Elite Scientists Sponsorship Program by Tianjin. One of the authors (G.D. Liu) acknowledges the financial support from Hebei Province Program for Top Young Talents.

**Figures and captions:**

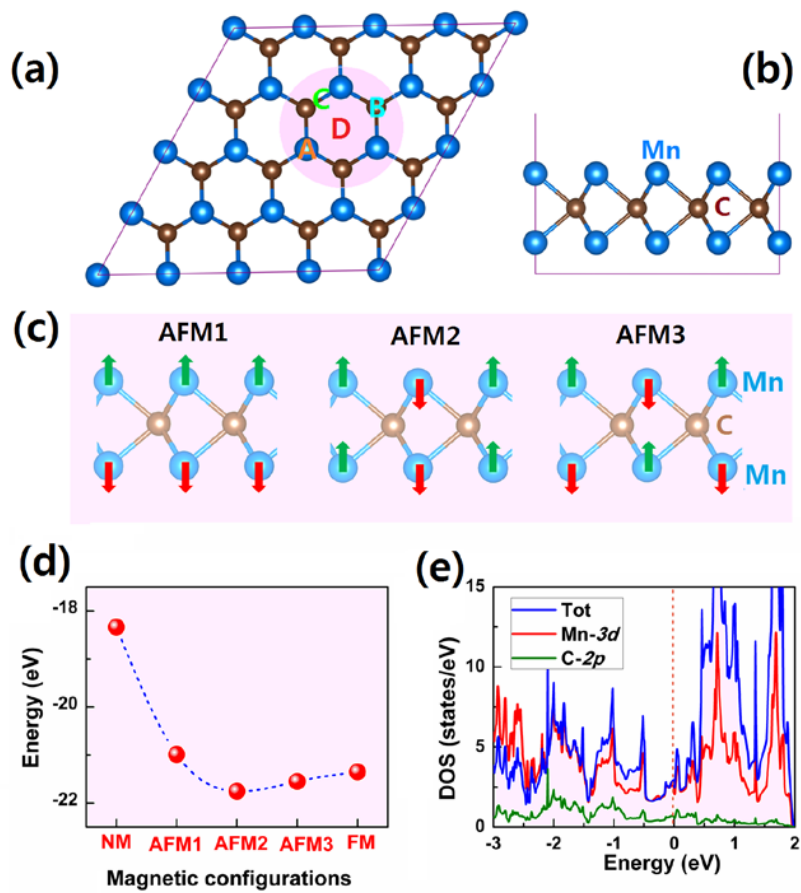

**Fig. 1** (a) Top and (b) side views of the atomic structures for $Mn_2C$ monolayer. In (a), sites A-D denote the possible H/O adsorption sites on $Mn_2C$ surface. (c) Three possible antiferromagnetic configurations (AFM1, AFM2, and AFM3) in $Mn_2C$ sheet. Blue and red arrows represent spin up and spin down, respectively. (d) Comparison of energies among different magnetic states including nonmagnetic (NM), ferromagnetic (FM), and antiferromagnetic (AFM1, AFM2, and AFM3). (e) Total DOS and partial DOS of $Mn_2C$ monolayer under AFM2 state (here only the DOS in spin up is shown).

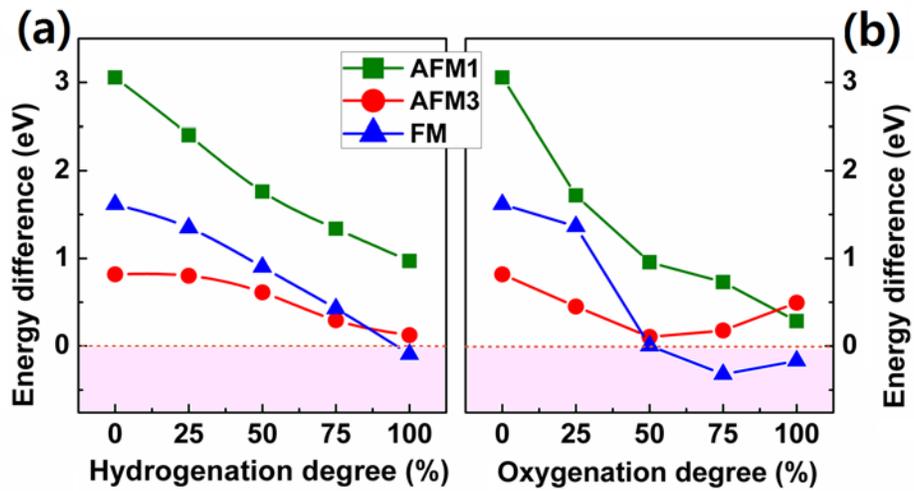

**Fig. 2** Comparison of energies among possible magnetic states for different (a) hydrogenation and (b) oxygenation degrees. In (a) and (b), the energy for the AFM2 state has been set as the reference energy. Note, the NM state has far higher energy than other magnetic states, and the energy for the NM state is not shown in the figures.

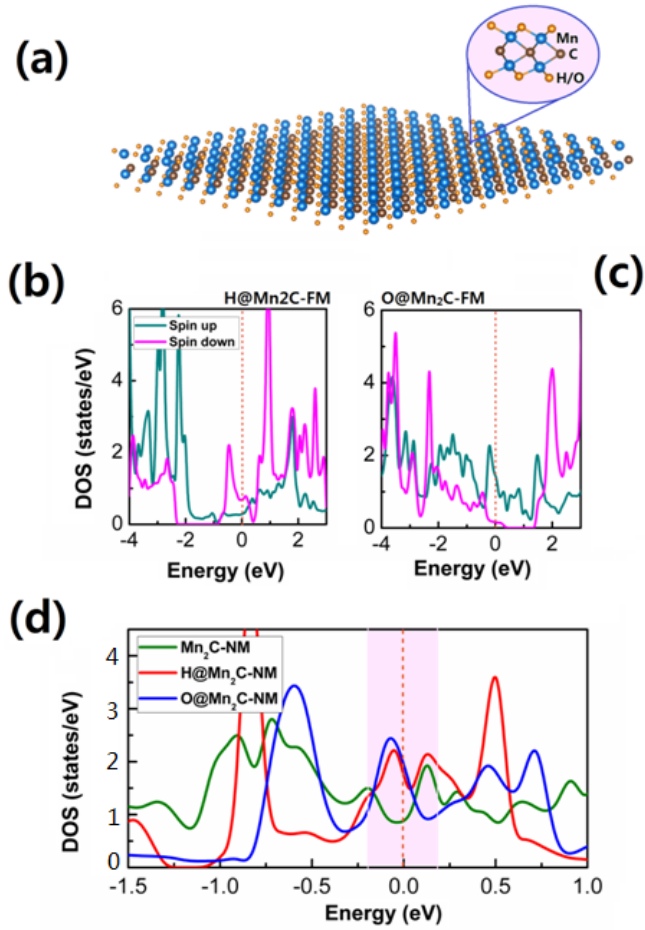

**Fig. 3** (a) The structural geometry for Mn$_2$C monolayer under fully hydrogenation/oxygenation. Total DOS for (b) hydrogenated and (c) oxygenated Mn$_2$C monolayer. (d) Comparison of total DOS under NM state for Mn$_2$C monolayer before and after hydrogenation/oxygenation.

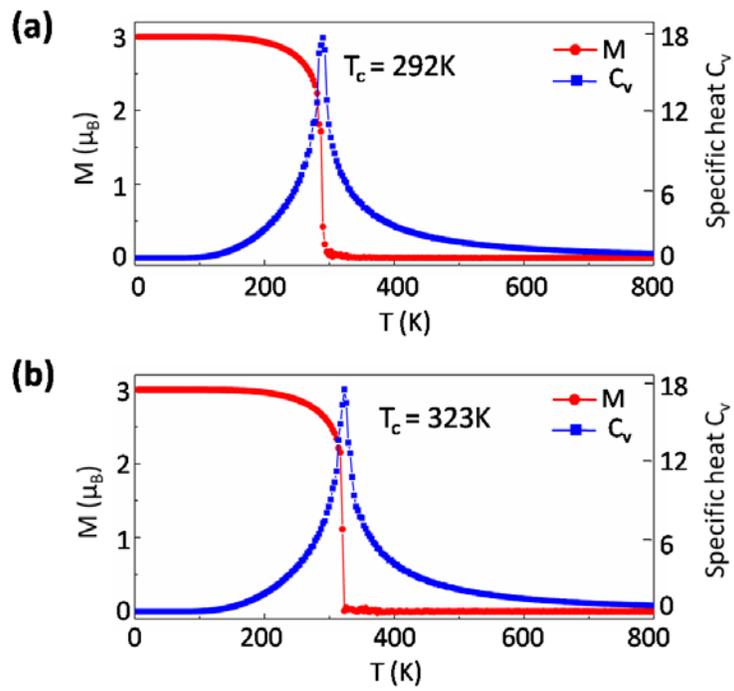

**Fig. 4** The simulated magnetic moment and specific heat as a function of temperature for (a) hydrogenated and (b) oxygenated $Mn_2C$ monolayer.

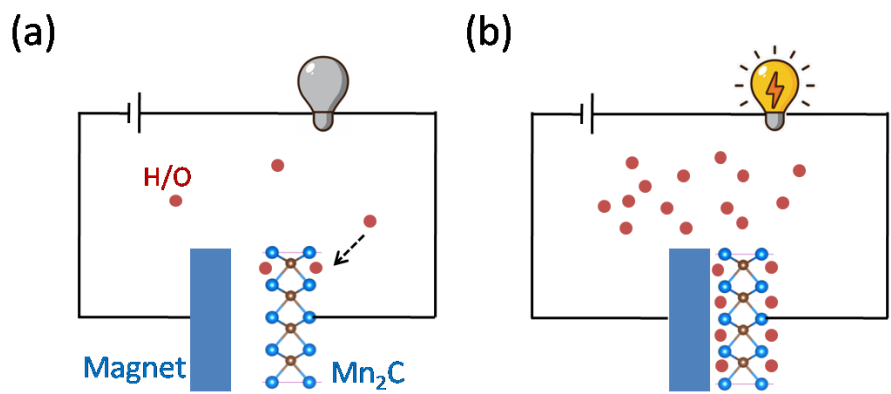

**Fig. 5** The simplified schematic diagrams of a Mn$_2$C-based H/O detector device. In (a), Mn$_2$C monolayer is in the AFM state under none or a low H/O atmosphere. On this occasion, the circuit is open. In (b), Mn$_2$C monolayer transfers into the FM state after fully hydrogenation/oxygenation. The circuit is connected, and the power indicator is lighted. The red dots in (a) and (b) denote H/O.

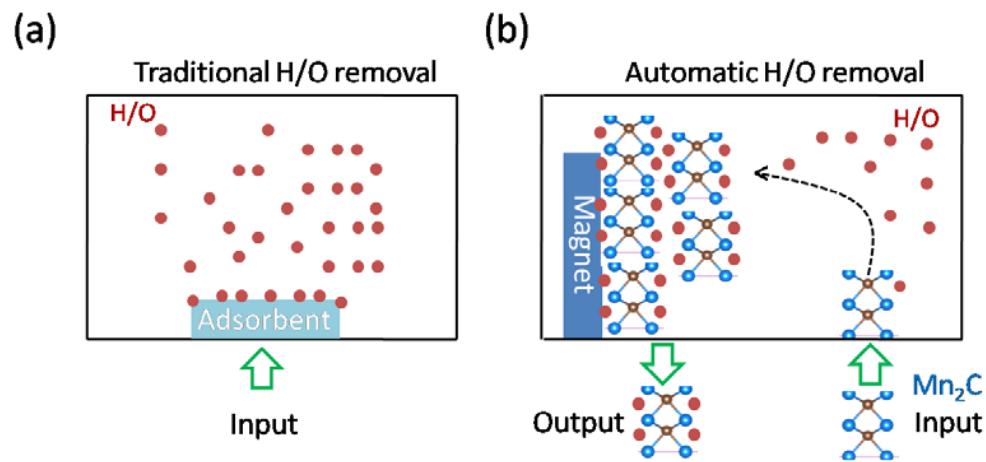

**Fig. 6** (a) Illustration of the mechanism for a traditional H/O removal device. The black rectangle represents the cavity containing H/O atmosphere. The red dot denotes H/O. The adsorbent is displayed for H/O adsorption. (b) is similar with (a), bur for the $Mn_2C$-based H/O removal device. In (b), the adsorbent ($Mn_2C$ monolayer) can be automatically caught by the magnet after reaching the maximum H/O adsorption capacity.